# Tuning the magnetic exchange via a control of orbital hybridization in $Cr_2(Te_{1-x}W_x)O_6$

M. Zhu[1], D. Do[1], C.R. Dela Cruz[2], Z. Dun[3], H. D. Zhou[3,4], S. D. Mahanti[1], and X. Ke[1,*]

[1.] *Department of Physics and Astronomy, Michigan State University, East Lansing, MI 48824, USA.*

[2.] *Quantum Condensed Matter Division, Oak Ridge National Laboratory, Oak Ridge, TN 37831, USA*

[3.] *Department of Physics and Astronomy, University of Tennessee, Knoxville, TN 37996, USA*

[4.] *National High Magnetic Field Laboratory, Florida State University, Tallahassee, FL 32306, USA.*

We report the complex magnetic phase diagram and electronic structure of $Cr_2(Te_{1-x}W_x)O_6$ systems. While compounds with different $x$ values possess the same crystal structure, they display different magnetic structures below and above $x_c = 0.7$, where both the transition temperature $T_N$ and sublattice magnetization ($M_s$) reach a minimum. Unlike many known cases where magnetic interactions are controlled either by injection of charge carriers or by structural distortion induced via chemical doping, in the present case it is achieved by tuning the orbital hybridization between Cr $3d$ and O $2p$ orbitals through W $5d$ states. The result is supported by *ab-initio* electronic structure calculations. Through this concept, we introduce a new approach to tune magnetic and electronic properties via chemical doping.

Competition between nearest- and further- neighbor magnetic interactions or competing interactions within special frustrated lattice structures often leads to a wide range of unusual magnetic ground states, which has been attracting intense attention in the community of condensed matter physics for the past few decades [1]. Relative strengths of these interactions can be readily tuned by external parameters, *i.e*, pressure, chemical doping, *etc*. Generally, non-isovalent chemical impurities substituted into nonmagnetic sites can introduce charge carriers which change the interaction pathway, such as Sr substitution on La in $La_{1-x}Sr_xMnO_3$ that induces a ferromagnetic double-exchange interaction [2]. On the other hand, isovalent impurities can give rise to various extents of lattice distortion due to different ionic radii, which in turn modulates the exchange couplings, as seen in $Sr_{2-2x}Ca_{2x}RuO_4$ [3]. In this work, we study the magnetic properties of $Cr_2(Te_{1-x}W_x)O_6$ system where isovalent W is substituted into Te sites. The presence of low-lying W $5d$ states and their hybridization with O 2p states that provide the exchange path between Cr $3d$ moments give an additional control on the nature and strength of this exchange. The resultant ferromagnetic exchange competes with the existing antiferromagnetic super-exchange interaction, leading to a complex magnetic phase diagram. This, we argue, introduces a new approach of tuning magnetic interactions through the control of orbital hybridization, which is different from the conventional charge carrier effect or pressure effect induced by chemical doping.

Magnetic properties of $Cr_2TeO_6$ (CrTO) and $Cr_2WO_6$ (CrWO) were first reported by Kunnmann *et al* in the late 1960s [4]. Both are insulators and undergo antiferromagnetic transitions at $T_N$ ~ 93 K and 45 K, respectively [5]. Intriguingly, despite being isostructural (both are ordered inverse-trirutile structures with tetragonal symmetry $P4_2/mnm$ as shown schematically in Fig. 1(a); lattice parameters $a_0$ = 4.545 Å and $c_0$ = 8.995 Å for CrTeO and $a_0$ =

4.583 Å and $c_0$ = 8.853 Å for CrWO) as well as very similar Cr-O-Cr bond angles and bond lengths [4,6] due to the similar ionic radii of $Te^{6+}$ and $W^{6+}$ ions, these two compounds display different magnetic structures [4] as shown in Fig. 1(b) and Fig. 1(c). One can visualize the different spin structures in terms of $Cr^{3+}$ bilayers, with each $Cr^{3+}$ bilayer separated by a (Te,W) layer. While in both cases the $Cr^{3+}$ spins are ferromagnetically aligned within each basal (*ab*) plane and antiferromagnetically coupled between the bilayers (inter-bilayer, *i.e*, Cr1 - Cr3, and Cr2 - Cr4), the neighboring intra-bilayer $Cr^{3+}$ spins (Cr1 - Cr2 and Cr3 - Cr4) are anferromagnetically aligned in CrTO (Fig. 1(b)) and ferromagnetically aligned in CrWO (Fig. 1(c)). It is important to emphasize that this difference in magnetic structure cannot be explained by the simple Goodenough-Kanamori rule [7,8] based on superexchange interactions, and thus other exchange mechanisms have to be invoked. In addition, the electronic structures in these systems have remained unexplored although the initial study of the compounds dates back to more than four decades. Furthermore, how the magnetic structure and the nature of the exchange interaction change in $Cr_2(Te_{1-x}W_x)O_6$ as one goes from Te to W are fundamental questions that need to be addressed.

In this Letter, we report the magnetic phase diagram and the electronic structure of $Cr_2(Te_{1-x}W_x)O_6$ system obtained via neutron powder diffraction and first-principles electronic structure calculations, respectively. We observe a crossover of magnetic structure in $Cr_2(Te_{1-x}W_x)O_6$ system at $x_c$ = 0.7, where both the transition temperature $T_N$ and sublattice magnetization ($M_s$) reach a minimum. Our *ab-initio* electronic structure calculations show that the presence of low lying W 5*d* states and their hybridization with O *p* states play an important role in the exchange interaction between Cr 3*d* moments, which can explain the different ground state spin configurations observed in the end compounds.

Figure 2(a) gives the lattice parameters at $T = 4$ K for $Cr_2(Te_{1-x}W_x)O_6$ [6], which are obtained by refining the neutron powder diffraction (NPD) data using FULLPROF program [9]. The in-plane lattice expands while the *c*-axis contracts slightly with increasing W doping, and the linear relationship of *a* and *c* as a function of *x* suggests a homogenous solid solution of Te and W mixture. The temperature dependence of magnetic susceptibility measured with an applied field of 3000 Oe is shown in Fig. 2(b). There are two noteworthy features: i) all samples display a broad peak with the peak position $T_p$ evolving non-monotonically with *x*; ii) as to be shown later, antiferromagnetic transition temperature $T_N$ of these samples is smaller than the corresponding $T_p$ of each sample, implying the development of strong low-dimensional correlation prior to forming 3D long-range magnetic order. The increase in susceptibility at low temperatures presumably originates from the paramagnetic impurities.

The non-monotonic dependence of $T_N$ on *x* is corroborated by the temperature dependence of magnetic heat capacity shown in Fig. 2(c). The magnetic heat capacity is obtained after subtracting the phonon contribution from the total heat capacity using the scaled heat capacity of an isostructural nonmagnetic compound, $Ga_2TeO_6$. Compared with the relatively sharp drop in heat capacity for $T > T_N$ for both the end members, the magnetic heat capacity of $Cr_2(Te_{1-x}W_x)O_6$ with non-zero *x*, particularly for $0.3 \leq x \leq 0.8$, exhibits broad peaks, reminiscent of a spin-glass like transition. However, no difference is found in the temperature dependence of *DC* magnetic susceptibility measured under field-cooled and zero-field-cooled conditions and the *AC* magnetic susceptibility measurements (data not shown) reveal nearly frequency ($f = 10 - 10$ kHz) independent signal even for $x = 0.5$ at which the compound shows the strongest chemical site disorder. This excludes the occurrence of a spin-glass transition. Instead, all the compounds undergo a long-range antiferromagnetic ordering as revealed by NPD measurements discussed next. On the other hand, together with the fact that the integrated saturated magnetic entropy is in the range of 4.43 - 9.53 J / K mol$_{Cr}$, smaller than the theoretical value (11.5 J / K mol$_{Cr}$) for $Cr^{3+}$ ions with $S = 3/2$, the broadening of the magnetic heat capacity implies the existence of strong magnetic fluctuations due to competing magnetic interactions. This suggests that in $Cr_2(Te_{1-x}W_x)O_6$ the low temperature long range ordered antiferromagnetic state coexists with a fluctuating paramagnetic background, a feature similar to $Tb_{2+x}Ti_{2-2x}Nb_xO_7$ [10].

Some representative NPD data for $x = 0$, 0.5, 0.8, and 1.0 measured at $T = 4$ K (and at 150 K shown in the insets) are shown in Fig. 1S in the Supplementary Material [6]. For $x = 0$ and 0.5, the magnetic Bragg peaks show up with a propagation wave vector $\boldsymbol{Q} = (0\ 0\ 2)$, and the refined spin structure is with antiferromagnetic intra-bilayers (AFM-I) as shown in Fig. 1(b); for $x = 0.8$ and 1.0, the magnetic propagation wave vector is $\boldsymbol{Q} = (0\ 0\ 1)$, and the corresponding magnetic structure is with ferromagnetic intra-bilayers (AFM-II) as shown in Fig. 1(c). It is noteworthy that the full-width at half-maximum (FWHM) of magnetic Bragg peaks of all samples is determined by the instrumental resolution according to the Rietveld refinement, confirming the existence of a long-range magnetic order at low temperatures. Temperature dependence of the order parameter for these four samples is plotted in Fig. 2(d) where one can see the non-monotonic dependence of $T_N$ on $x$, as discussed previously, with $T_N$ for $x = 0.8$ smaller than the others.

The $T_N$ - $x$ phase diagram of $Cr_2(Te_{1-x}W_x)O_6$ is shown in Fig. 1(d). Interestingly, this system displays a crossover of magnetic state at $x_c = 0.7$. The compounds with $x < 0.7$ exhibit an AFM-I type magnetic structure, while the compounds with $x > 0.7$ show an AFM-II type magnetic structure. Accordingly, $T_N$ varies non-monotonically as a function of $x$ and reaches a minimum value ($T_N \sim 29.6$ K) at $x = 0.7$. At the crossover point $x_c = 0.7$, one sees a coexistence of both AFM-I and AFM-II types of magnetic structures [6]. Intriguingly, the magnetization $M$ of $Cr_2(Te_{1-x}W_x)O_6$ obtained from the data refinement at $T = 4$ K also displays a non-monotonic dependence on $x$ and reaches a minimum at $x_c$, as shown in Fig. 1(e). All $M$ values are much smaller than the expected ideal theoretical value $M_s = 3.0$ $\mu_B$ for $Cr^{3+}$ ions (with negligible spin-orbit coupling).

The above experimental observations bring out several interesting and related questions: i) What are the underlying mechanisms determining the magnetic ground states of the end members? Because of very similar crystal structures and lattice parameters of CrTO and CrWO, the difference in their magnetic structures cannot be explained simply by the Goodenough-Kanamori rule based on superexchange interactions. Thus, one has to understand the differences in the electronic structures of these materials. ii) Why do both $T_N$ and $M$ depend non-monotonically on $x$ and display minimum values at $x_c = 0.7$?

In order to understand the ground state magnetic structure and the nature of intra- and inter-bilayer exchange interactions, we have carried out density functional theory (DFT) calculations [6] within generalized gradient approximation (GGA) and GGA+U [11] as implemented in the Vienna ab-initio Simulation Package (VASP) [11,12,13], using projector-augmented wave (PAW) method [14,15] and Perdew-Burke-Ernzerhof (PBE) exchange-correlation functional [16]. We have chosen 4 different long-range ordered magnetic states denoted as A-B (AF - AF = AFM-I, AF - F = AFM-II, F - AF and F - F), where A refers to inter-bilayer and B refers to intra-bilayer magnetic ordering. In all calculations, structural parameters and ionic positions are allowed to relax.

In Table 1a we give the GGA energies (per magnetic unit cell containing four Cr atoms). The ground state is AF-AF (AFM-I) for the Te compound and AF-F (AFM-II) for the W compound in agreement with the neutron diffraction results. The magnetic moments are nearly the same for all the four Cr atoms and lie in the range 2.6 - 2.8 $\mu_B$, lower than the value of 3.0 $\mu_B$ for the $Cr^{3+}$ spin, indicating hybridization between Cr $d$ and O $p$, Te $s$, and W $d$ states. Since GGA generally does not adequately describe the $d$-electrons of transition metal atoms, we have also done GGA+U calculations [17,18]. In the same Table we give the energies for U = 4 eV (this incorporates both intra-site Coulomb repulsion and exchange through a single parameter [11]). The lowest energy states are consistent with those obtained in GGA calculations. The major effect of U is to reduce the splitting between the ground and excited states indicating a reduction in the strength of effective exchange coupling between the Cr moments. At the same time the magnetic moments of Cr atoms increase to ~ 3.0 $\mu_B$.

To understand the nature of the exchange couplings between different Cr moments, we look at the geometry and local coordinations of inter- and intra- bilayer nearest neighbor (nn) Cr pairs. The distance between the inter-bilayer nearest-neighbor (nn) Cr atoms (Cr1 and Cr3) is ~ 3.00 Å whereas the distance between intra-bilayer nn Cr atoms (Cr1 and Cr2 or Cr3 and Cr4) is ~ 3.60 Å. The exchange interaction between Cr1 and Cr3 is dominated by Anderson antiferromagnetic (A-AFM) kinetic exchange [19] (~ $2t^2/U$ in the Hubbard model representation where $t$ is the hopping between the $d$-orbitals of Cr and U is an effecting intra-atomic Coulomb repulsion). This direct A-AFM exchange between intra-bilayer nn Cr atoms (Cr1 and Cr2) is likely to be negligible because the distance is ~ 3.60 Å and $t$ falls off exponentially. On the other hand, for both intra- and inter-dimers, one has to consider the superexchange via O (SE-O) atom which is bonded to either a Te or a W atom. Since Cr3-O-Cr1 angle is close to 90$^o$, we expect the strength of Cr1-Cr3 SE-O to be weak, thus A-AFM exchange dominates leading to an antiferromagnetic alignment. In contrast, Cr1-O-Cr2 (or Cr3-O-Cr4) angle is ~ 130$^o$ and SE-O should be appreciable. For CrTO this SE-O is antiferromagnetic, consistent with filled oxygen states providing the SE path. On the other hand, for CrWO the Cr1-O-Cr2 coupling is ferromagnetic (FM) whose origin can be attributed to the low-lying unoccupied W-$d$ state that hybridizes with the O-$p$ that mediates this exchange. One plausible scenario is that W-$d$ and O-$p$ hybridization creates a virtual hole (either spin up or down) which mediates a ferromagnetic double exchange between the two Cr moments [6]. This underlying physics is somewhat similar to one proposed by Kasuya to explain the ferromagnetic coupling between rare-earth moments in EuS [20].

In Fig. 3, we give the ground state spin densities for the two compounds. The spin densities associated with Cr1 and Cr3 (inter-bilayer coupling) and the nearest oxygen (O1) are very similar in the two compounds. Also the superexchange through these oxygen atoms (SE-O) is very small due to near $90^o$ exchange path (as one can see in Fig. 3), and different O $p$ orbitals hybridize with Cr1 and Cr3. In contrast, the spin densities associated with intra-bilayer exchange between Cr1 and Cr2 (or Cr3 and Cr4) differ dramatically between the two compounds. For Te this exchange is dominated by O2 induced superexchange ($130^o$ path) with very little Te $s$ or $p$ state mixing, whereas for W the O2 charge and spin distributions are strongly altered by the W $d$ states. This hybridization (and the basic difference between Te and W compounds) is also seen in the projected density of states given in the supplementary material [6]. In Table 1b we give the values of inter- (J) and intra- (j) bilayer exchange obtained by fitting the energies of different spin configurations obtained within GGA to a spin 3/2 Heisenberg model. Clearly the nn inter-bilayer antiferromagnetic exchange (J) is dominant in both, the intrabilayer exchange is antiferromagnetic for CrTeO and ferromagnetic for CrWO. Introduction of the intrasite Coulomb repulsion U reduces the strengths of the exchange, particularly the antiferromagnetic exchange (Table 1(b)). These results are in qualitative agreement with the values extracted from high temperature susceptibility measurements by Drillon et al [21], although there are quantitative differences (see Ref. 6 for further discussions).

Finally, to understand the magnetic structure of $Cr_2(Te_xW_{1-x})O_6$ for different $x$, we have used the virtual crystal approximation (VCA) and GGA to calculate the energies of the AFM-I and AFM-II structures. The energy difference between antiferromagnetic and ferromagnetic intra-bilayers is plotted in Fig 4. We find that the ground state switches from AFM-I to AFM-II when $x \sim 0.7$, consistent with the experimental results. Although VCA addresses the problem in an average way (it does not probe the effects of local fluctuations caused by disorder, clustering, etc.), the results suggest that by controlling the effective coupling between Cr $3d$ and W $5d$ states (indirectly through intervening oxygen) through the substitution of W into Te sites, we can indeed tune the competition of the magnetic interactions of the intra-bilayers, i.e, AFM SE-O and FM induced by the orbital hybridization. At $x \sim 0.7$ these two magnetic interactions are comparable in strength, leading to the strongest spin fluctuation and thus the minimum values in both $M_s$ and $T_N$.

In summary, we have discovered an unexpected crossover of the magnetic ground states at $x_c \sim 0.7$ in isostructural $Cr_2Te_{1-x}W_xO_6$ compounds where both $T_N$ and $M_s$ reach a minimum. We have attributed these phenomena to competing magnetic interactions that originate from an unusual contribution to magnetic exchange that comes from orbital hybridization between low energy unoccupied W $5d$ substituted into Te sites and O $2p$ states which provide the exchange path between two Cr moments. This work highlights a new approach to tune the magnetic exchange via chemical doping without introducing additional charge carriers or structural distortion.

X.K. acknowledges the support from the start-up funds at Michigan State University. Z.D. and H. D. Z. thank for the support from JDRD program of University of Tennessee. D.D and S.M. acknowledge the support by the Center for Revolutionary Materials for Solid State Energy Conversion, an Energy Frontier Research Center funded by the DOE Office of Basic Energy Sciences under Award No. DE-SC0001054. Work at ORNL was supported by the Scientific User Facilities Division, Office of Basic Energy Sciences, DOE.

**Figure Captions**

Figure 1. (a) Schematic of the bilayer crystal structure of $Cr_2(Te_{1-x}W_x)O_6$. Each $Cr^{3+}$ bilayer is separated by a (Te,W) layer. (b,c) Schematics of two different types of antiferromagnetic spin structures: AFM-I (b) with spins antiferromagnetically aligned within the bilayer and AFM-II (c) with spins ferromagnetically aligned within the bilayer. Spins of neighboring bilayers are antiferromagnetically coupled for both AFM-I and AFM-II. (d) $T_N$ - $x$ phase diagram of $Cr_2(Te_{1-x}W_x)O_6$. PM represents the paramagnetic phase. (e) Magnetization as a function of $x$ obtained from neutron powder diffraction measurements.

Figure 2. (a) $x$ dependence of lattice parameters measured at $T = 4$ K. (b) and (c) show the temperature dependence of magnetic susceptibility and magnetic heat capacity, respectively. (d) Normalized neutron scattering intensity of ordering parameters, (1 0 1) for $x = 0$, 0.5 and (0 0 1) for $x = 0.8$, 1.0, as a function of temperature. Solid curves are the guide to the eyes.

Figure 3. Spin densities (on (110) plane) of $Cr_2TeO_6$ (left) and $Cr_2WO_6$ (right) where red: spin up, blue: spin down.

Figure 4. Energy difference between AF and F intra-bilayer magnetic configurations with AF configuration fixed for the inter-bilayer (AF) interaction.

Table 1. (a) Total energy (eV/unit cell) of different magnetic configurations for [Inter]-[Intra] bilayer. AF: Antiferomagnetic, F: Feromagnetic. (b) Calculated and experimental values of inter- (J) and intra- (j) bilayer exchange parameters as in the model proposed by Drillon *et al*. [21] using GGA.

(a)

| Configuration | | AF-AF | AF-F | F-AF | F-F |
|---|---|---|---|---|---|
| GGA | $Cr_2TeO_6$ | -131.711 | -131.547 | -131.634 | -131.411 |
| | $Cr_2WO_6$ | -159.930 | -160.002 | -159.854 | -159.815 |
| GGA+U, U=4 | $Cr_2TeO_6$ | -122.603 | -122.572 | -122.583 | -122.561 |
| | $Cr_2WO_6$ | -150.714 | -150.768 | -150.701 | -150.750 |

(b)

| Compound | | $Cr_2TeO_6$ | | $Cr_2WO_6$ | |
|---|---|---|---|---|---|
| Parameter | | J (meV) | j (meV) | J (meV) | j (meV) |
| Theo. | GGA | -4.3 | -2.3 | -10.4 | 1.0 |
| | GGA+U, U=4 eV | -1.12 | -0.46 | -1.0 | 0.75 |
| Ref. [21] | | -2.9 | -0.4 | -3.8 | 0.12 |

Figure 1.

M. Zhu *et al*

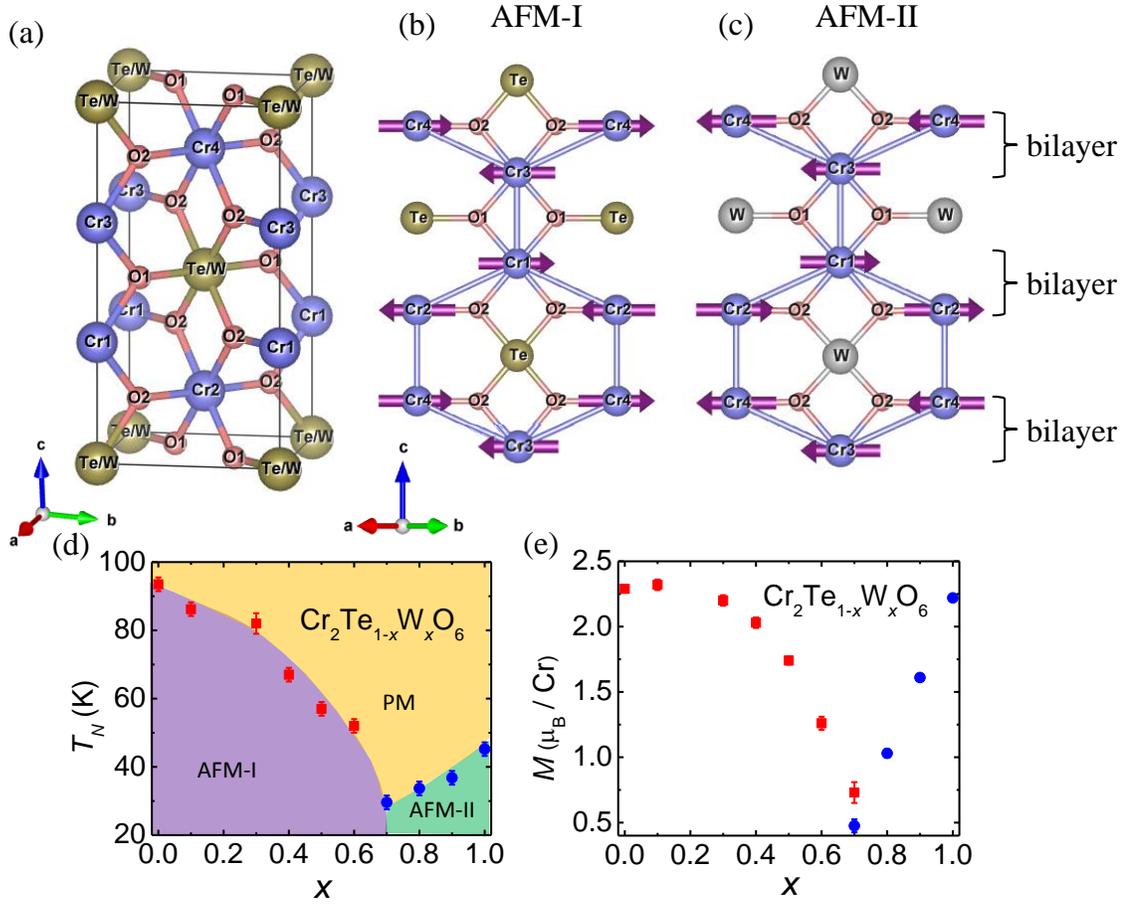

Figure 2.

M. Zhu *et al*

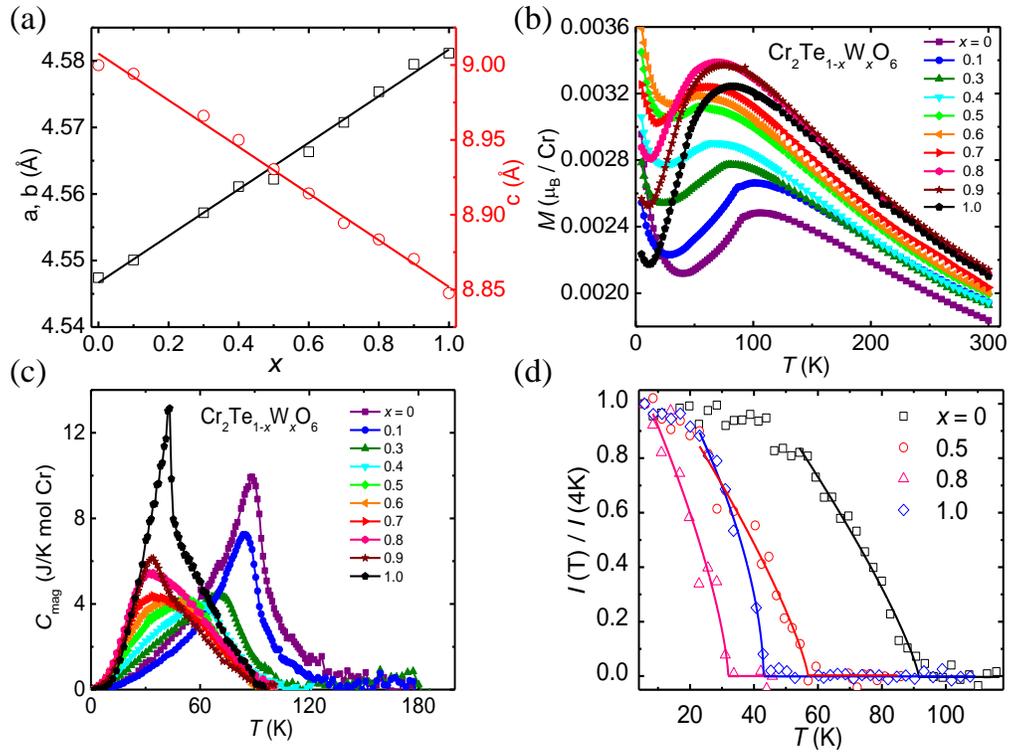

Figure 3.

M. Zhu *et al*

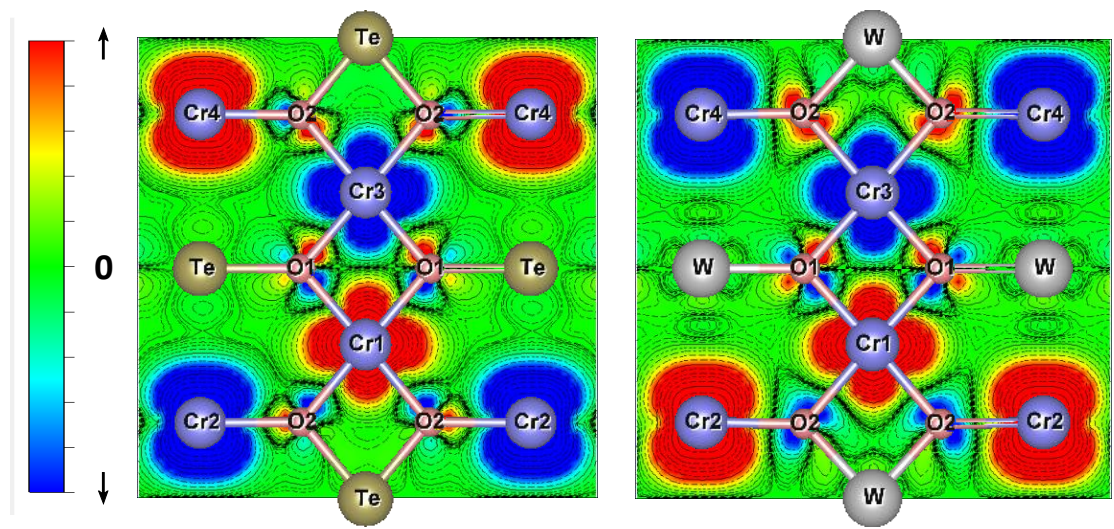

Figure 4

M. Zhu *et al*

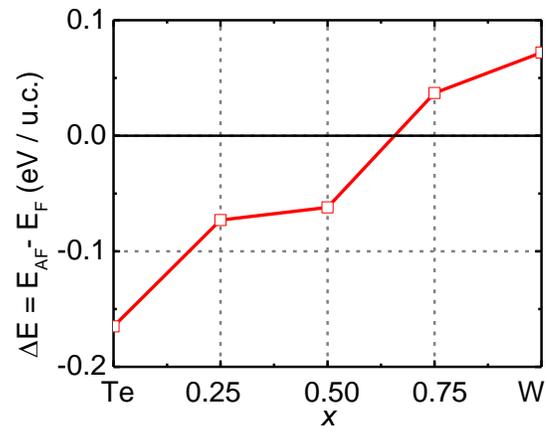